\RequirePackage{fix-cm}
\documentclass[smallextended]{svjour3}       
\smartqed  

\usepackage{graphicx}
\usepackage{mathptmx}      
\usepackage{latexsym}
\usepackage{amsmath}
\usepackage{subfig}
\usepackage{xcolor}
\usepackage{epstopdf}
\begin{document}

\title{Experimental Implementation of a NMR NOON State Thermomete}

\author{
C. V. H. B. Uhlig, R. S. Sarthour, I.S. Oliveira \and
A. M. Souza}

\institute{C. V. H. B. Uhlig \at
              Centro Brasileiro de Pesquisas Fisicas \\
              Tel.: +55 21 2141-7362\\
              \email{cyntiavhb@gmail.com}       
           \and
           R. S. Sarthour \at
              Centro Brasileiro de Pesquisas Fisicas
			   \and
           I.S. Oliveira \at
              Centro Brasileiro de Pesquisas Fisicas
			   \and
               A. M. Souza \at
              Centro Brasileiro de Pesquisas Fisicas
}

\date{Received: date / Accepted: date}

\maketitle

\begin{abstract}
Utilizing the highly correlated quantum NOON states of particles, we have implemented a proof-of-principle quantum thermometer using the NMR technique for measuring the variation of local magnetic field with the temperature variation. The system used was the star-topology system of hexafluorophosphate molecules and the thermometer showed a sensitivity of $85nT/ ^o C$.
Using the hexafluorophosphate and the trimethylphosphite spin systems, we have quantified the advantage of the quantum protocol over the classical one for measuring magnetic field. The quantum protocol showed the best performance for a time evolution in the quantum circuit of $T = 20ms$, where the errors in the measurement scaled as the Heisenberg limit $1/N$.
\keywords{Quantum metrology \and Quantum thermometry \and NOON states \and NMR Quantum Computing}
\end{abstract}

\section{Introduction}
\label{intro}

Quantum metrology \cite{degen} is the field of study of performing high precise measurement using quantum phenomena, usually exploring quantum correlations {\cite{giovannetti1}-\cite{girolami}}. Typically, a quantum object is initialized in an suitable state and then let to evolve for some period of time. During this period, the quantum probe acquires information about the quantity of interest $\mathcal{Q}$. The value of such quantity is then estimated by measuring the probe. Repeating the process $N$ times with independent probes, the minimal uncertainty achivable related to the estimation of $\mathcal{Q}$ scales with $1/\sqrt{N}$, the well known Standard Quantum Limit (SQL). However this limit is not the ultimate limit to the precision, but can be beated when the probes are quantum correlated. Using special entangled states, for example, the uncertainty scales with $1/N$, the Heisenberg Limit, as demonstrated in some recent experiments {\cite{israel}-\cite{napolitano}}, {\cite{10spin}-\cite{shukla}}.

Quantum themometry is a special issue of quantum metrology that deals with the ultimate limit to the precision at which the temperature of a system can be determined \cite{stace}. In standard thermometers, the temperature is inferred measuring a physical property of the thermometer in thermal equlibrium with the system of interest. In quantum thermometry, the thermometer does not need to thermalize with the target system, instead, a quantum thermometer is a quantum object in an superposition state that encodes the temperature in the relative phase between its quantum states. 

A non-thermalizing thermometer that uses an interferometric approach and quantum correlations was experimentaly demonstrated in a Nuclear Magnetic Resonance (NMR) setup \cite{cesar} and experimental implementations of nanoscale quantum thermometers were also reported in recent works {\cite{tham}-\cite{highpre}}. However, up to our knowledge, there is no demonstration of a quantum thermometer that works at the Heisenberg limit. In the present paper we have used  the NMR technique to implement a proof-of-principle quantum thermometer based on a interferometric approach and entangled states.  The thermometer is an ensemble of nuclear spins prepered in the so called NOON state \cite{dowling}.

\section{NMR Themometer in Ensamble Quantum Metrology}
\label{sec:1}

The useful states for quantum metrology are those that contain quantum correlations \cite{giovannetti2}, such as NOON states \cite{modi}, {\cite{highnoon}-\cite{khurana}} and  squeezed states \cite{sqzd}. In a optical Mach-Zehnder setup, for example, the NOON state is a superposition of all photons traveling through a channel A and all photons passing through a channel B.  In our experiment we considered a NMR setup analogous to a Mach-Zehnder experiment where the quantum probes are nuclear spins in a large ensemble of non-interacting molecules.  Considering $N$  spins, the NOON state is a superposition of all spins in the ``spin-up" state  ($|0\rangle$)  and all spins in ``spin-down" state  ($|1\rangle$ )

\begin{equation}
| \psi \rangle = \frac{| 000 ... \rangle + | 111 ... \rangle }{\sqrt{2}}.
\label{noon}
\end{equation} 

On the other hand, a single spin, initially prepared in the state $(|0\rangle + |1\rangle)/\sqrt{2}$  precess in the presence of a magnetic field $\delta$,  so that after the time interval $\Delta t$ it envolves to the state $(|0\rangle + e^{\gamma \delta \Delta t}|1\rangle)/\sqrt{2}$ where $\gamma$ is the gyromagnetic ratio of the nucleus. This phenomenon can be used to build NMR magnetic field sensors {\cite{10spin}-\cite{simmons}} in which a local magnetic field is detected as a phase shift observed in the NMR spectrum.  It is interesting to note that the shift observed in the spectrum is analogous to the interference pattern observed in optical experiments.  An important difference between the present case and optical setups is that all probes are collectively and continuously measured at same time (ensemble average).

In a standard NMR sensor, the spins envolve independently, however, if the spins are initially prepared in a correlated state, for example the NOON state,  then an enhanced sensitivity to the field can be obtained \cite{10spin}. The set of molecules suitable for prepering NOON states are those with the ``star topology'', in which a nuclear spin $A$ interacts with $N$ magnetically equivalent spins $X$ through $J$ coupling, forming a system of $N+1$ nuclear spins. In the NMR terminology those systems are denominated as $AX_{N}$ systems, the methyl $^{13}CH_3$ and methylene $^{13}CH_2$ groups, commonly found in organic molecules, are exemples of $AX_{3}$ and $AX_{2}$ star topology systems, respectively. Protocols for magnetic field sensing based on quantum metrology were demonstrated using the trimethylphosphite (TMP) molecule \cite{10spin} , comprised by a central phosphorus atom and nine magnetically equivalent hydrogens, and the tetramethylsilane (TMS) \cite{simmons}, which has a central silicon atom surrounded by twelve hydrogens. The star-topology systems are also demonstrably useful for other applications, for instance, as quantum registers using a TMS system and a 37-spin system \cite{pande}.

The resonance frequency of a particular nucleus is determined by the local field it senses. The local magnetic field, in turn, depends upon the chemical nature of the molecule in which they reside. The difference in the local magnetic field caused by the chemical environment is called chemical shift \cite{levitt} and may depend on the temperature \cite{RKHarris}. In this paper we have implemented a NMR quantum thermometer using NOON states to monitor the variation of the local magnetic field caused by chemical shifts variation.

\section{Experimental Procedure}
\label{sec:2}

\begin{figure}[h!]
\centering
\subfloat{\includegraphics[width=0.7\textwidth]{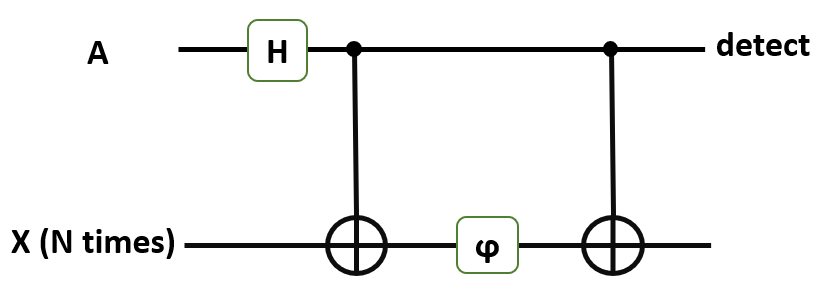}}
\subfloat{\includegraphics[scale=0.4]{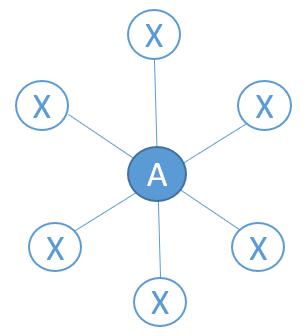}}
\caption{Left: Quantum circuit representing the quantum metrology protocol for measurring magnetic field. Right: Diagram of a molecule $A X_{N}$ with star topology symmetry.}
\label{RES:01}
\end{figure}

\begin{figure}[h!]
\includegraphics[width=1.1\textwidth]{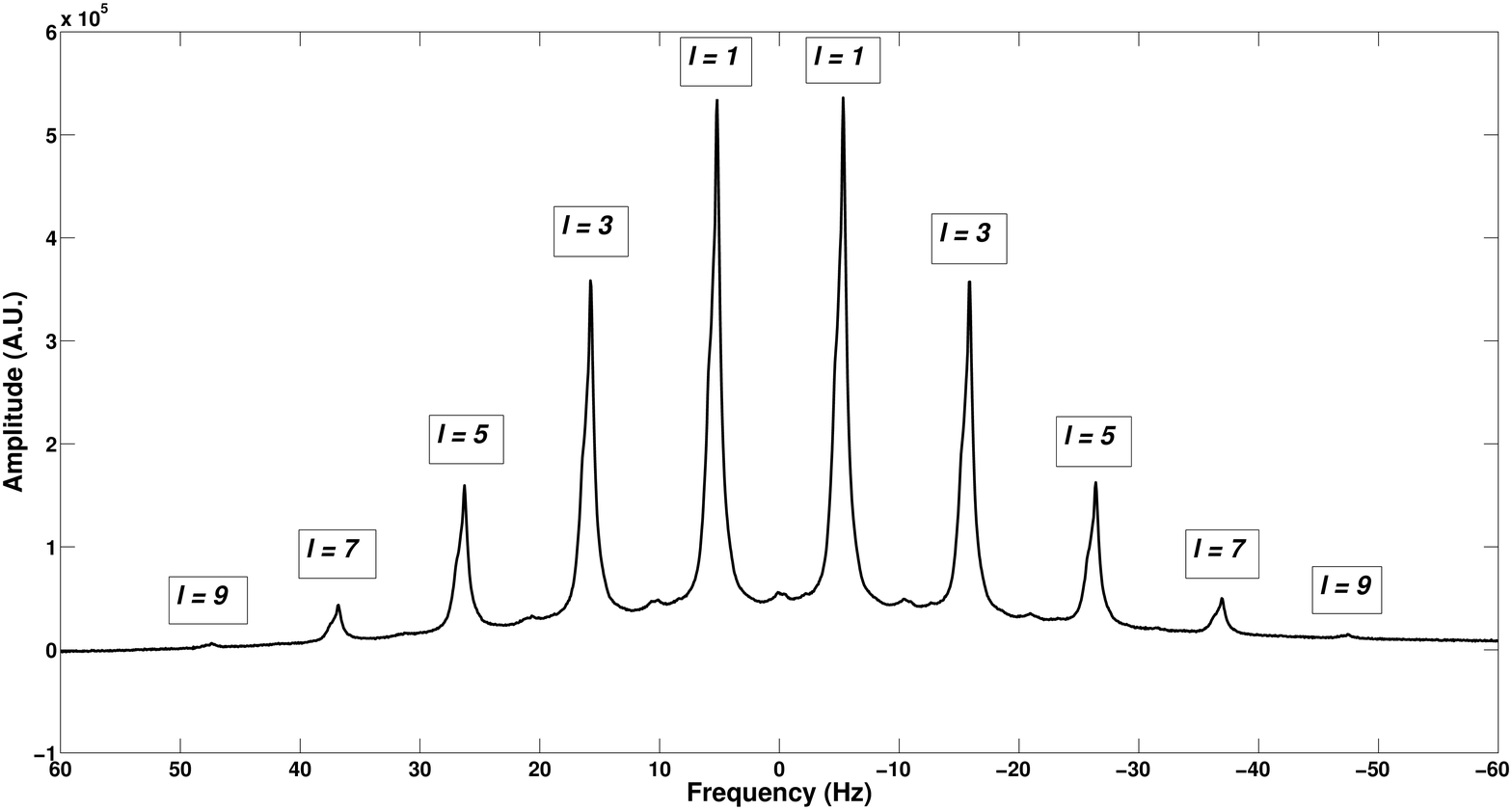}
\caption{Phosphorus central spin spectrum for the TMP sample. The outermost peaks, around frequencies 50 Hz and -50 Hz, are hardly visible due to the low polarization of the NOON states ($l=9$) on the sample.}
\label{RES:02}
\end{figure}

\begin{figure}[h!]
\includegraphics[width=1.1\textwidth]{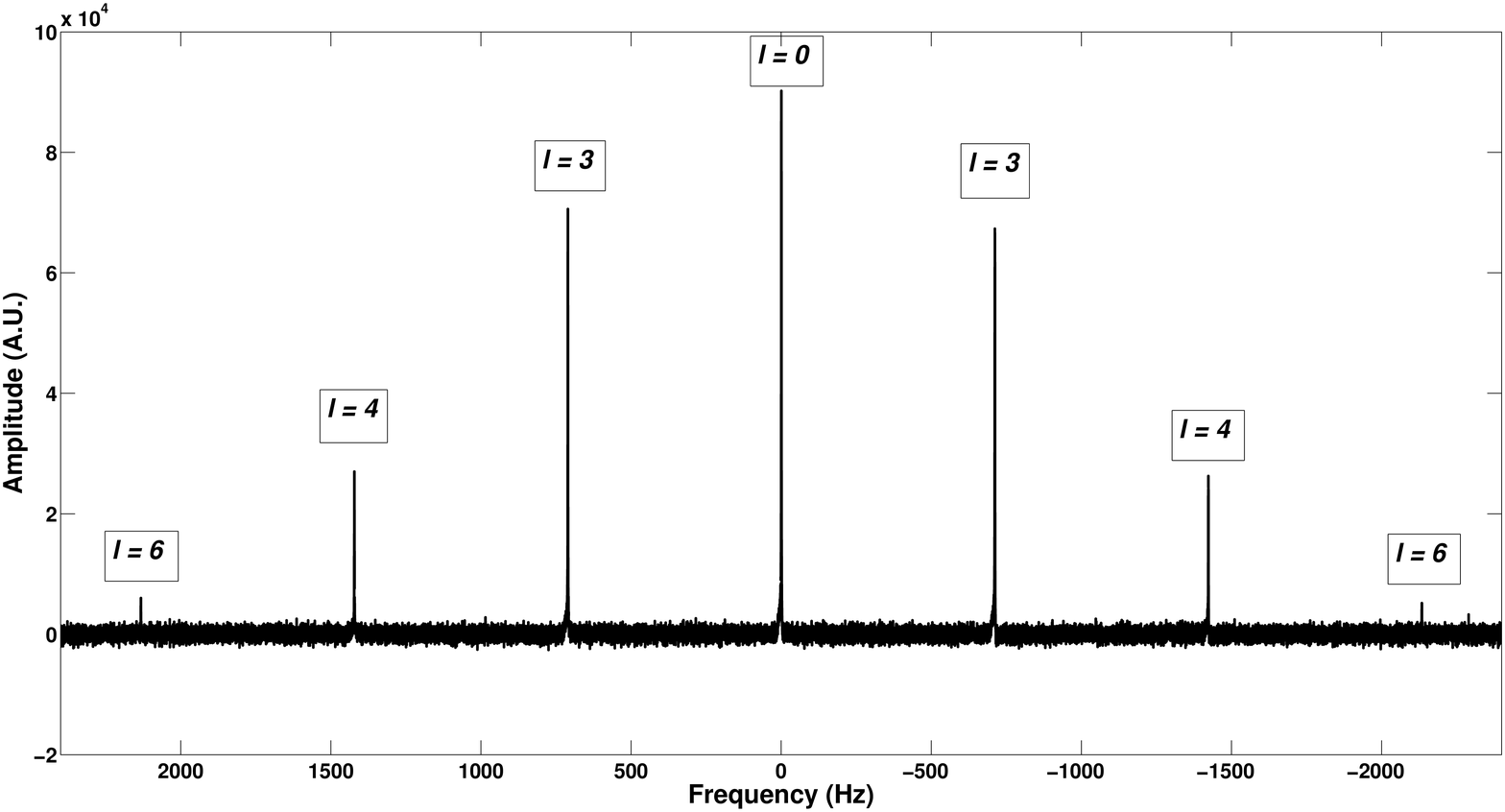}
\caption{Phosphorus central spin spectrum for the hexafluorophosphate sample. In this sample the outermost peaks, around frequencies 2100 Hz and -2100 Hz, are more visible than for the TMP sample, here the NOON states corresponding to the $l= 6$.}
\label{RES:03}
\end{figure}

The quantum circuit used in the experiment is shown in figure (\ref{RES:01}). The first part of the circuit consists of a Hadamard gate applied on the central spin $A$ followed by a CNOT gate on the target satellite spins. Due to the symmetry of the star topology, the CNOT gate acts on all spins at same time. When the system is initialized in the state $|000 ... \rangle$, the circuit creates the NOON state (\ref{noon}). After the state preparation, the system is let to envolve freely for a period $T$, this period is used to encode the temperature in the relative phase between the quantum states of eq. (\ref{noon}). At the end of the protocol a new CNOT gate is needed for the readout. The temperature is inferred observing the phase shift of the central spin spectra. In order to enhance the signal to noise ratio of the central spins,  before the NOON state preparation the polarization of the central spin in enhanced by applying a NMR sequence known as Insensitive Nuclei Enhanced by Polarization Transfer (INEPT), not shown in (\ref{RES:02}) and (\ref{RES:03}).  All gates were implemented using standard NMR-Quantum information processing techniques \cite{ivan}.  

It is important to mention that the experiment is performed at room temperature, therefore our spin system is initialized at thermal equilibrium.  In this situation, the output of the first part of the circuit (figure (\ref{RES:01}) ) is not the pure state eq. (\ref{noon}), but the statistical mixture

\begin{equation}
\rho_ = \sum_{l} \rho_{l} 
\label{roh}
\end{equation}

where

\begin{equation}
\rho_{l} = \sum _{k=0}^{k_l} |U,D \rangle \langle U,D| 
\label{sud}
\end{equation}
and $l = |U-D|$. $U$ is the number of spins-up and $D$ is the number of spins down, when $l = N$ then the sum in eq. (\ref{sud}) will have only one term corresponding to the NOON state (eq. (\ref{noon})). If $l \neq N$ there will be $k_l$ indistinguishable permutations of $|U,D \rangle$ states. All states with the same $l$  aquires phase at same rate. Different $l$ states can be distinguished observing the appropriate resonance line of the central spins \cite{10spin}.  In figures (\ref{RES:02}) and (\ref{RES:03}) we show the correspondence of the central spin resonance lines and each possible $l$ state for the two molecules used in this work, the TMP and hexafluorophosphate molecules.

\section{Experimental Results}

\begin{figure}[h!]
\includegraphics[width=1.1\textwidth]{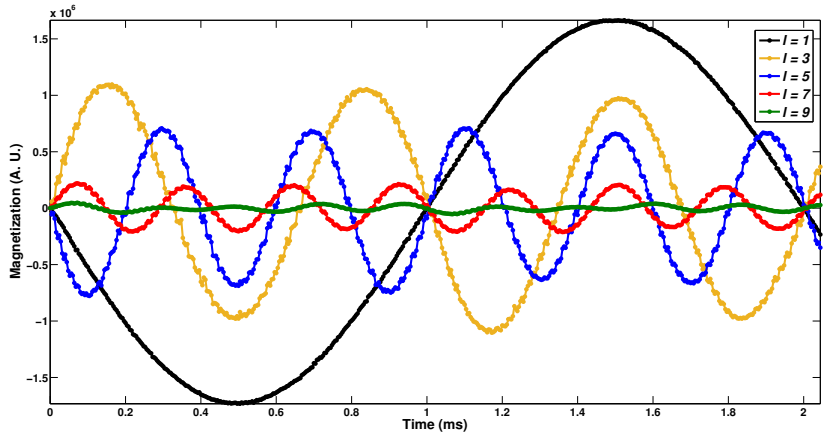}
\caption{TMP spectrum peaks evolution, for $T_{max}=2ms$. The smallest amplitude curve corresponds to the state with $l=9$. The greater the amplitude the smaller the corresponding $l$ ($l=9, 7, 5, 3, 5 , 1$).}
\label{RES:04}
\end{figure}

\begin{figure}[h!]
\includegraphics[width=1.1\textwidth]{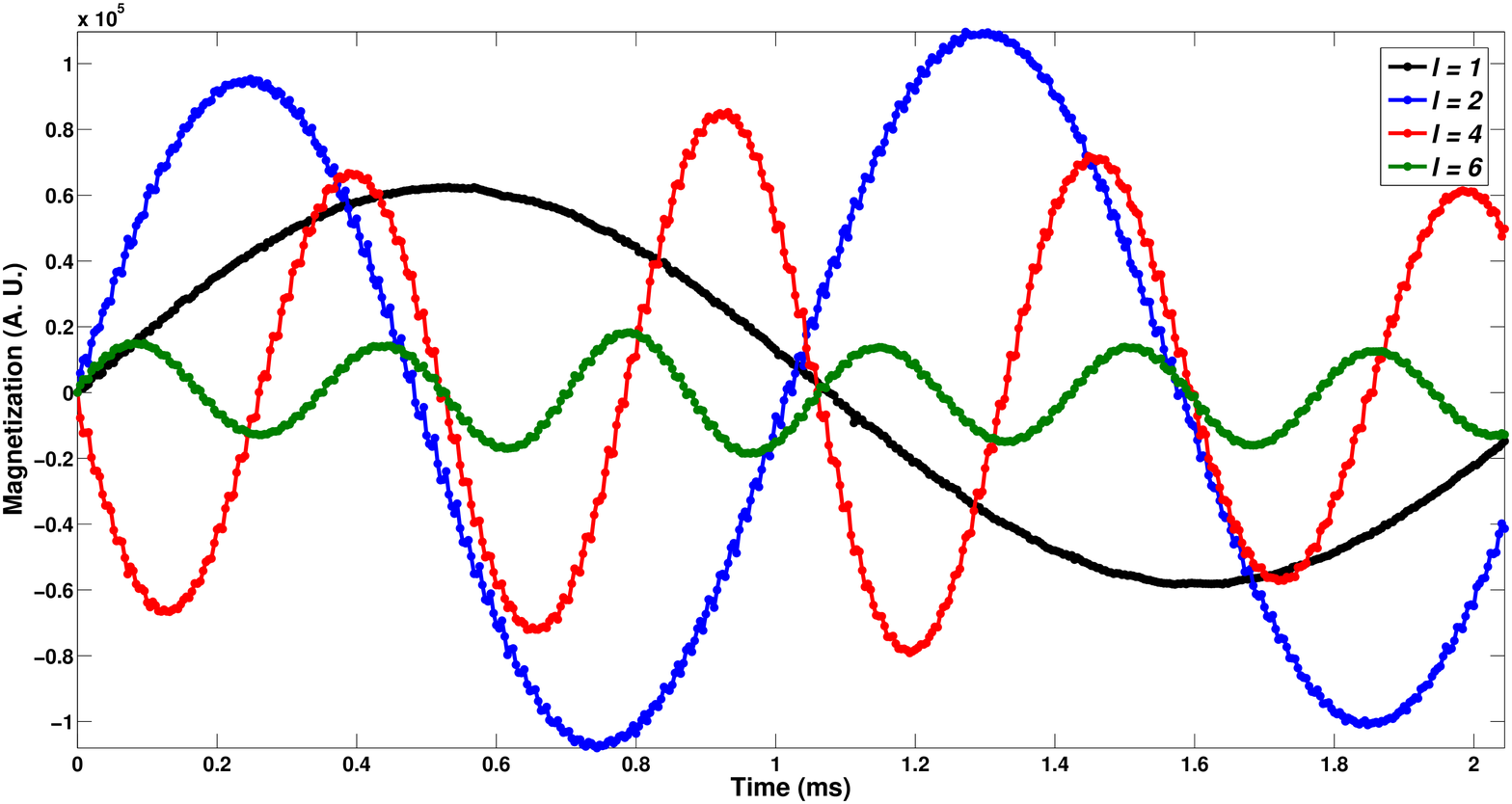}
\caption{Hexafluorophosphate spectrum peaks evolution, for $T_{max}=2ms$. 
The smallest amplitude curve corresponds to the state with $l=6$. The greater the amplitude the smaller the corresponding $l$ ($l=6, 4, 2, 1$).}
\label{RES:05}
\end{figure}

The performance of quantum metrology protocols are limited by decoherence effects. Before the implementation of the quantum thermometer we have performed a series of experiments to study the influence of decoherence effects. The experiments were performed using a Varian 500 MHz spectrometer. In the experiments, an offset field of $\delta= 11.73\mu T$ was set and the protocol given in section (\ref{sec:2}) was applied. The evolution of each resonance line of the central spin spectrum observed for TMP and hexafluorophosphate samples are shown in figures (\ref{RES:04}) and (\ref{RES:05}). A total of 512 spectra were acquired, one for each time value of evolution $T_{max}$. It was verified that the outermost peaks oscillated faster than the inner ones, that is, they acquired phase faster in response to the field, showing a higher sensitivity to it, as can be seen in figures (\ref{RES:04}) and (\ref{RES:05}). The oscillations with the smallest amplitudes and the highest frequencies refer to the states with $l = 9$, for the TMP sample, and with $l = 6$, for the hexafluorophosphate sample. The larger amplitude oscillation in figure (\ref{RES:04}) refers to isolated (uncoupled) hydrogen nuclei, $l = 1$, and in figure (\ref{RES:05}), refers to the isolated fluorine nuclei, also with $l = 1$. 

As a next step, we performed Fourier transforms with respect to the time $t$. The transformed spectrum were fitted in order to estimate the offset field and the error associated to the measurement. 

\begin{figure}[h!]
\includegraphics[width=1.1\textwidth]{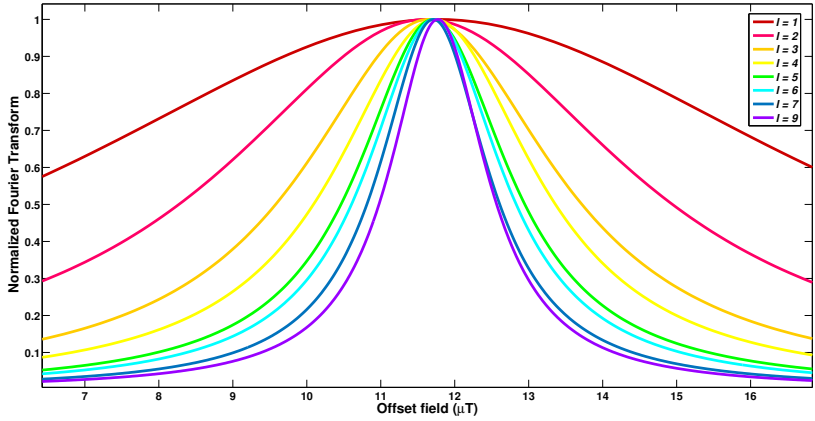}
\caption{Normalized Fourier Transform with respect to the time $t$ for all states of the TMP and the hexafluorophosphate sample, for $T_{max}=2ms$. From inward to outward the curves correspond to $l=9,7,6,5,4,3,2,1$ states. It can be noticed that all the peaks are positioned around the predicted value of the magnetic field (an offset of $\delta = 11.73mT$). The estimation of the magnetic field uncertainty falls as states with higher $l$ are used.}
\label{RES:06}
\end{figure}

The fits provided the results shown in figure (\ref{RES:06}), for both samples, with number $l$ odd corresponding to the TMP sample and the even valeus of $l$ to the hexafluorophosphate sample. 

As we can see, states with high $l$ values has smaller error associated, where the error is quantified as Full Width at Half Maximum (FWHM) of the fitted data. 
We have quantified the advantage of the quantum protocol by the ratio between the error obtained for the classical protocol, where the spin evolves uncorrelated ($l=1$), and the error obtained for the quantum strategy. 

\begin{figure}[h!]
\includegraphics[width=1.1\textwidth]{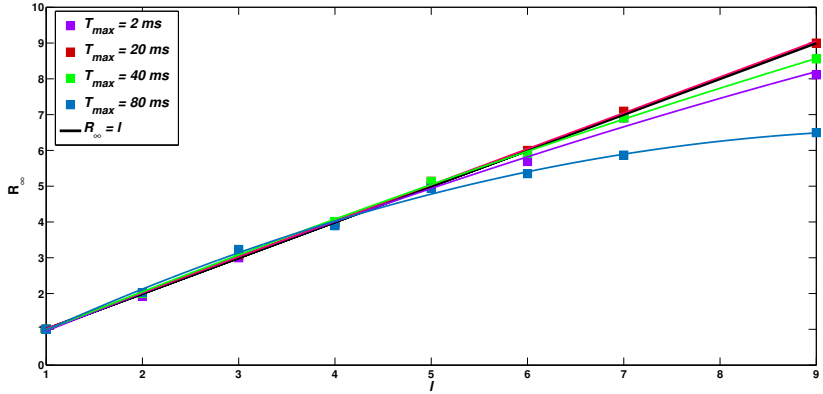}
\caption{Ratio of the isolated spin error (``classical error'') over NOON state error ($R_{\infty}$) \textit{versus} $l$. The amount $R_{\infty}$ refers to the advantage of whether or not to use a quantum strategy of preparation of states (following \cite{schaffry})}
\label{RES:07}
\end{figure}

The performance of the quantum protocol is shown in figure (\ref{RES:07}). For small values of $l$ the quantity $R_{\infty}$ scales linearly with $N$, which indicates that the uncertainty of the parameter estimation scales as $1/N$, analogous to the Heisenberg limit. As $l$ increases, the observed performance of the quantum protocol is worse then expected, due to the fact that decoherence effects are much more severe for high order NOON states. However, in all cases showed in the figure (\ref{RES:07}), the quantum protocol is advantageous over the classical counterpart. The best performance occurs for $T_{max} = 20 ms$; the existence of an optimal time was predicted in ref. \cite{schaffry}. 

\begin{figure}[h!]
\includegraphics[width=1.1\textwidth]{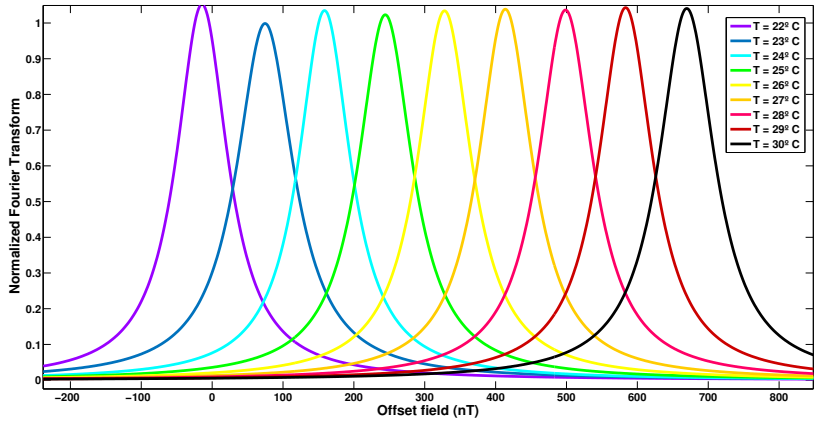}
\caption{Normalized Fourier Transform with respect to time $t$ for the NOON state ($l=6$) of the hexafluorophosphate sample measured at $9$ values of temperature, with $T_{max} = 53ms$. The figure shows the increase of chemical shift according to the increase in the value of the temperature in the sample, from few nanoteslas of magnetic field to hundreds of nanotesla of the field.}
\label{RES:08}

\end{figure}

\begin{figure}[h!]
\includegraphics[width=1.1\textwidth]{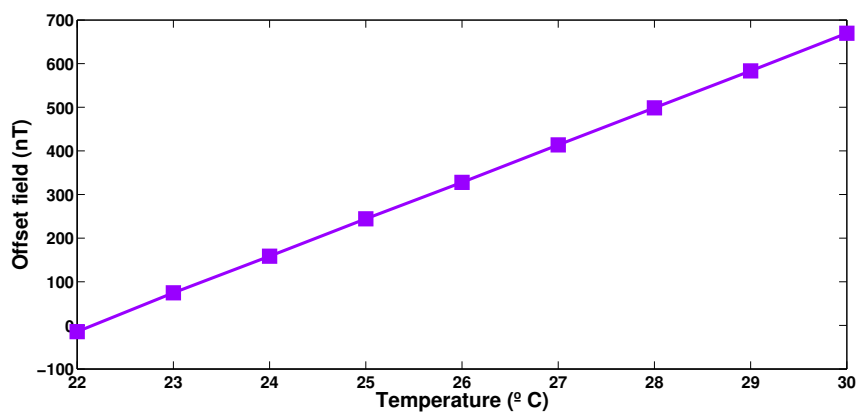}
\caption{Offset field (chemical shift) \textit{versus} temperature for the $NOON$ states of the hexafluorophosphate sample with $N=6$.}
\label{RES:09}
\end{figure}

The calibration of the thermometer was performed fixing the temperature at $22 ^oC$ and  $30 ^oC$. For each temperature the quantum protocol (figure (\ref{RES:01}) ) was applied to measure the chemical shift variation due to the temperature change. From the difference between the chemical shifts at $22 ^oC$ and  $30 ^oC$ we have calibrated the thermometer and found that the sensitivity of the thermometer is $\approx 85  nT/ ^o C $. After that, the temperature was changed from  $22 ^oC$ and  $30 ^oC$ at steps of $1^oC$, Fourier transforms with respect to $T$ were performed (figure (\ref{RES:08})) for the NOON state ($l=6$) of the hexafluorophosphate sample and provided the chemical shift as function of the temperature shown in figure (\ref{RES:09}).


\section{Conclusion}

In summary, we have implemented a proof-of-principle NOON state thermometer that using the NMR technique. The experiment was performed in ensemble of a large number of independent molecules in liquid state. As far as we know this is the first experimental demonstration of a NOON state thermometer. Quantum thermometry can have many applications. Future works will be devoted to study the availability to develop NMR applications of quantum themometry, such as monitoring the temperature and the confection of temperature maps \cite{tmap} of samples during chemical reactions. Works will also be devoted to investigate ways to improve quantum thermometry protocols by using dynamical decoupling methods \cite{alexandre}, \cite{khurana}, \cite{dorfman} to mitigate decoherence effects. 
 
We acknowledge financial support from the Brazilian agency, CAPES and CNPq. This work was performed as part of the Brazilian Na-
tional Institute of Science and Technology (INCT) for Quantum Information Grant No. 465469/2014-0.  AMS  acknowledges support from FAPERJ (203.166/2017).



\end{document}